\documentclass[10pt, conference, compsocconf]{IEEEtran}

\usepackage{cite}

\usepackage[dvips]{graphicx}

\usepackage[cmex10]{amsmath}
\usepackage[caption=false,font=footnotesize]{subfig}

\usepackage{url}

\usepackage[dvipdfm,CJKbookmarks=true,bookmarksopen=true,colorlinks,citecolor=black,linkcolor=black,anchorcolor=black,urlcolor=black] {hyperref}
\usepackage{balance}

\usepackage{mathptmx}
\usepackage{amsfonts, multirow, amssymb}

\usepackage[linesnumbered, ruled]{algorithm2e}

\begin{document}

\title{To Vote Before Decide: A Logless One-Phase Commit Protocol\\for Highly-Available Datastores}

\author{\IEEEauthorblockN{Yuqing Zhu{\small $~^{\#1}$}, Philip S. Yu{\small $~^{\vartriangle2}$}, Guolei Yi{\small $~^{+3}$}, Wenlong Ma{\small $~^{\#4}$}, Mengying Guo{\small $~^{\#4}$}, Jianxun Liu{\small $~^{\#4}$}}
\IEEEauthorblockA{$~^{\#}$ICT, Chinese Academy of Sciences, Beijing, China\\
$~^{\vartriangle}$University of Illinois at Chicago, USA\\
$~^{+}$Baidu, Bejing, China\\
$~^{1}$zhuyuqing@ict.ac.cn, $~^{2}$psyu@uic.edu, $~^{3}$yiguolei@baidu.com\\
$~^{4}$\{mawenlong,guomengying,liujianxun\}@ict.ac.cn}
}


\maketitle

\begin{abstract}
Highly-available datastores are widely deployed for online applications. However, many online applications are not contented with the simple data access interface currently provided by highly-available datastores. Distributed transaction support is demanded by applications such as large-scale online payment used by Alipay or Paypal. Current solutions to distributed transaction can spend more than half of the whole transaction processing time in distributed commit. An efficient atomic commit protocol is highly desirable. This paper presents the \textbf{HACommit} protocol, a logless one-phase commit protocol for highly-available systems. HACommit has transaction participants vote for a commit before the client decides to commit or abort the transaction; in comparison, the state-of-the-art practice for distributed commit is to have the client decide before participants vote. The change enables the removal of both the participant logging and the coordinator logging steps in the distributed commit process; it also makes possible that, after the client initiates the transaction commit, the transaction data is visible to other transactions within one communication roundtrip time (i.e., one phase). In the evaluation with extensive experiments, HACommit outperforms recent atomic commit solutions for highly-available datastores under different workloads. In the best case, HACommit can commit in one fifth of the time 2PC does.
\end{abstract}


\IEEEpeerreviewmaketitle

\section{Introduction}

Online applications have strong requirements on availability; their data storage widely exploits highly-available datastores~\cite{f1,amazonDown,facebookMemcache}. For highly-available datastores, distributed transaction support is highly desirable. It can simplify application development and facilitate large-scale online transacting business like Paypal~\cite{paypal}, Alipay~\cite{alipay} or Baidu Wallet~\cite{bwallet}. Besides, it can enable quick responses to big data queries through materialized view and incremental processing~\cite{percolator,view}. The benefits of transactions come from the ACID (atomicity, consistency, isolation and durability) guarantees~\cite{concurrencybook}. The atomic commit process is key to the guarantee of ACID properties. Current solutions to atomic commit incurs a high cost inhibiting online applications from using distributed transactions. A fast atomic commit process is highly desirable.

The state-of-the-art practice for distributed commit in highly-available datastores is to have the transaction client decide before the transaction participants vote~\cite{tapir,helios,spanner,scatter,mdcc,rcommit}, denoted as the \emph{vote-after-decide} approach. On deciding, the client initiates a distributed commit process, which typically incurs two phases of processing. Participants vote on the decision in the first phase of commit, with the votes recording in logs or through replication. The second phase is for notifying the commit outcome and applying transaction changes. Even if the transaction client can be notified of the commit outcome at the end of the first phase~\cite{tapir,helios,mdcc,rcommit}, the commit is not completed and the transaction result is not visible to other transactions until the end of the second phase. The two processing phases involve at least two communication roundtrips, as well as the step for logging to write-ahead logs~\cite{rstar}  or for replicating among servers. The communication roundtrips and the logging or replicating step are costly procedures in distributed processing. They lead to a long distributed commit process, which then reduces transaction throughputs.

A different approach to distributed commit is having participants vote for a commit before the client decides to commit or abort the transaction, denoted as the \emph{vote-before-decide} approach. Having the participants vote first, the voting step can overlap with the processing of the last transaction operation, saving one communication roundtrip; and, the votes can be replicated at the mean time, instead of using a separate processing step. This makes the removal of one processing phase possible. On receiving the client's commit decision, the participants can directly commit the transaction locally; thus, the transaction data can be made visible to other transactions within one communication roundtrip time, i.e., one phase. Though previous one-phase commit protocols also have participants vote early, they need to make several impractical assumptions, e.g., log externalization~\cite{1pc}; besides, they rely heavily on the coordinator logs to guarantee atomicity and durability.

In this paper, we present the HACommit protocol, a logless one-phase commit protocol for highly-available systems. HACommit takes the vote-before-decide approach. In order to remove logging and enable one-phase commit, HACommit tackles two key challenges: the first is how to commit(abort) a transaction correctly  in a one-phase process; and, the second is how to guarantee a correct transaction recovery on participant or coordinator failures without using logs.

For the first challenge, we observe that, with the vote-before-decide approach, the commit process becomes a problem that the client proposes a decision to be accepted by participants. This problem is widely known as the consensus problem~\cite{misunderstanding}. Consensus algorithms are solutions to the consensus problem. The widely used consensus algorithm Paxos~\cite{paxos} can reach a consensus among participants (acceptors) in a one-phase process, if the proposer is the initial proposer in a run of the algorithm. HACommit runs the Paxos algorithm once for each transaction commit(abort). It uses the unique client as the initial proposer of the algorithm and the participants as the acceptors and learners. Thus, the client can propose any value, either commit or abort, to be accepted by participants as the consensus. HACommit proposes a new procedure for processing the last transaction operation such that consensus algorithms can be exploited in the commit process. To exploit Paxos, HACommit designs a transaction context structure to keep Paxos configuration information for the commit process.

For the second challenge, we notice that consensus algorithms can reach an agreement among a set of participants safely even on proposer failures. As HACommit exploits Paxos and uses the client as the proposer/coordinator, the client failure will not block the commit process. On the client failure, HACommit runs the classic Paxos algorithm to reach the same transaction outcome among the participants, which act as would-be proposers replacing the failed client. Furthermore, we observe that, in practice, the high availability of data in highly-available datastores leads to an equal effect of fail-free participants during commit. Instead of using logs for participant failure recovery, HACommit has participants replicate their votes and the transaction metadata to their replicas when processing the last transaction operation. For participant replica failures, HACommit proposes a recovery process that exploits the replicated votes and metadata.

With HACommit, a highly-available datastore can not only respond to the client commit request within one phase, as in other state-of-art commit solutions~\cite{tapir,mdcc,rcommit}, but also makes the transaction changes visible to other transactions within one phase, increasing the transaction concurrency.  Without client failures, HACommit can commit a transaction within two message delays. Based on Paxos, HACommit is non-blocking on client failures; and, it can also tolerate participant replica failures. HACommit can be used along with various concurrency control schemes~\cite{1pc,concurrencybook}, e.g., optimistic, multi-version or lock-based concurrency control.  We implemented HACommit  and evaluated its performance using a YCSB-based transaction benchmark~\cite{ycsb}. As the number of participants and data items involved in a transaction is the key factor affecting the performance of commit protocols, we evaluated HACommit and several recent protocols~\cite{spanner,rcommit,mdcc} by varying the number of operations per transaction. In the evaluation with extensive experiments, HACommit can commit in less than a millisecond. In the best case, HACommit can commit in one fifth of the time that the widely-used 2PC commits.\vspace{12pt}

\textbf{Roadmap}. Section~\ref{sec:related} discusses related work. Section~\ref{sec:overview} overviews the design of HACommit. Section~\ref{sec:lastop} details the last operation processing in HACommit and Section~\ref{sec:commit} describes the commit process. Section~\ref{sec:failure} presents the recovery processes on client and participant failures. We report our performance evaluations in Section~\ref{sec:eval}. The paper is brought to a close with conclusions in Section~\ref{sec:conclude}.

\section{Related Work}%
\label{sec:related}

\textbf{Atomic commit protocols} (ACPs). A large body of work studied the atomic commit problem in distributed environment both in database community~\cite{rstar,1pc,txnConsensus} and distributed computing community~\cite{md3pc,dnbac}. The most widely used atomic commit protocol is two-phase commit (2PC)~\cite{concurrencybook}. It has been proposed decades ago, but remains widely exploited in recent years~\cite{walter,rococo,hstore,rcommit,spanner}. 2PC involves at least two communication round trips between the transaction coordinator and the participants. Relying on both coordinator and participant logs for fault tolerance, it is blocking on coordinator failures.

Non-blocking atomic commit protocols were proposed to avoid the blocking on coordinator failures during commit. But some assume the impractical model of synchronous communication and incur high costs, so they are rarely implemented in real systems~\cite{3pc}. Those assuming the asynchronous system model generally exploit coordinator replication and the fault-tolerant consensus protocol~\cite{txnConsensus,md3pc}. These non-blocking ACPs generally incur an even higher cost than 2PC. Besides, they are all designed taking the same vote-after-decide approach as 2PC, i.e., that participants vote after the client decides.

One-phase commit (1PC) protocols were proposed to reduce the communication costs of 2PC. Compared to 2PC, they reduce both the number of forced log writes and communication roundtrips. The price is to send all participants' logs to the coordinator~\cite{1pccoordlog} or to make impractical assumptions on systems, e.g, consistency checking on each update~\cite{1pc}. Non-blocking 1PC protocols also exist. They have the same problems as blocking 1PC protocols. Though 1PC protocols have participants vote for commit before the client decides as HACommit does, they do not allow the client to abort the transaction if all transaction operations are successfully executed~\cite{concurrencybook}. In comparison, HACommit gives the client all the freedom to abort a transaction.

All the above atomic protocols do not consider the high availability of data as a condition, thus involving unnecessary logging steps for failure recovery at the participants or the coordinator. Exploiting the high availability of data, the participant logging step can be easily implemented as a process of data replication, which is executed for each operation in highly-available datastores--no matter the operation belongs to a transaction or not.

\textbf{ACPs for highly-available datastores}. In recent years, quite a few solutions are proposed for atomic commit in highly-available datastores. Spanner~\cite{spanner} layers two phase locking and 2PC over the non-blocking replica synchronization protocol of Paxos~\cite{paxos}. Spanner is non-blocking due to the replication of the coordinator's and participants' logs by Paxos, but it incurs a high cost in commit. Message futures~\cite{msgfutures} proposes a transaction manager that utilizes a replication log to check transaction conflicts and exchange transactions information across datacenters. The concurrency server for conflict checking is the bottleneck for scalability and performance. Besides, the assumption of shared logs are impractical in real systems~\cite{1pc}. Helios~\cite{helios} also exploits a log-based commit process. It can guarantee the minimum transaction conflict detection time across datacenters. However, it relies on a conflict detection protocol for optimistic concurrency control using replicated logs, which makes strong assumptions on one replica knowing all transactions of any other replica within a critical time interval, which is impossible  for asynchronous systems with disorderly messages~\cite{flp}. The safety property of Helios in guaranteeing serializability can be threatened by the fluctuation of cross-DC communication latencies. These commit proposals heavily exploit transaction logs, while logging is costly for transaction processing~\cite{perfbreakdown}.

MDCC~\cite{mdcc} proposes a commit protocol based on Paxos variants for optimistic concurrency control~\cite{concurrencybook}. MDCC exploits the application server as the proposer in Paxos, while the application server is in fact the transaction client. Though its application server can find out the transaction outcome within one processing phase, the commit process of MDCC is inherently two-phase, i.e., a voting phase followed by a decision-sending phase, and no concurrent accesses are permitted over outstanding options during the commit process. TAPIR~\cite{tapir} has a Paxos-based commit process similar to that of MDCC, but TAPIR can be used with pessimistic concurrency control mechanisms. It also uses the client as the proposer in Paxos. It layers transaction processing over the inconsistent replication of highly-available datastores, and exploits the high availability of data for participant replica recovery. TAPIR also returns the transaction outcome to the client within one processing phase of commit, but the transaction outcome is only visible to other transactions after two phases. It has strong requirements for applications, e.g., pairwise invariant checks and consensus operation result reverse. Replicated commit~\cite{rcommit} layers Paxos over 2PC. In essence, it replicates two-phase commit operations among datacenters and uses Paxos to reach consensus on the commit decision. It requires the full replica in each data center, which processes transactions independently and in a blocking manner.

All the above ACPs for highly-available datastores take the vote-after-decide approach. In comparison, HACommit exploits the vote-before-decide approach to enable the removal of one processing phase and the removal of logging in commit. HACommit overlaps the participant voting with the processing of the last operation. Using the unique client as the transaction coordinator and the initial Paxos proposer, HACommit commits the transaction in one phase, at the end of which the transaction data made visible to other transactions. HACommit exploits the high availability of data for failure recovery, instead of using the classic approach of logging.
\begin{figure}[!t]
      \centering
      \includegraphics[width=0.48\textwidth]{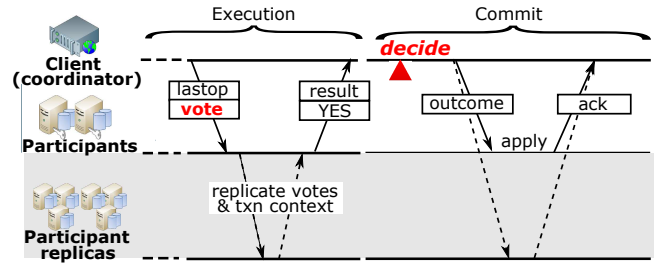}
      \caption{An example commit process using HACommit.}\vspace{-12pt}
      \label{fig:hacommit} 
\end{figure}
%
%
%
\section{Overview of HACommit}%
\label{sec:overview}

HACommit is designed to be used in highly-available datastores, which guarantee high availability of data. Generally, highly-available datastores partition data into shards and distribute them to networked servers to achieve high scalability. To guarantee high availability of data, each shard is replicated across a set of servers. Clients are front-end application servers or any proxy service acting for applications. Clients can communicate with servers of the highly-available datastores. A transaction is initiated by a client. A transaction participant is a server holding any shard operated by the transaction, while servers holding replicas of a shard are called participant replicas.

The implementation of HACommit involves both client and server sides. On the client side, it provides an atomic commit interface via a client-side library for transaction processing. On the server side, it specifies the processing of the last operation and the normal commit process, as well as the recovery process on client or participant failures. Except for the last operation, all transaction operations can be processed following either the inconsistent replication solutions~\cite{tapir,mdcc} or the consistent replication solutions~\cite{spanner,rcommit}. Different concurrency control schemes and isolation levels~\cite{concurrencybook} can be used with HACommit, e.g., optimistic, multi-version, lock-based concurrency control or read-committed, serializable isolation levels. On processing the last transaction operation, participants vote for a transaction commit based on the results of local concurrency control, integrity and consistency checks.

A HACommit application begins a transaction, starting the transaction execution phase. It can then execute reads and writes in the transaction. On the last operation, the client indicates to all participants that it is the last operation of the transaction. All participants check locally whether to vote YES or NO for a commit. They replicate their votes and the transaction context information to their replicas respectively before responding to the client. The client will receive votes for a commit from all participants, as well as the processing result for the last operation. This is the end of the execution phase. Then, the client can either commit or abort the transaction, though the client can only commit the transaction if all participants vote YES~\cite{graytxnbook}.

The atomic commit process starts when the client proposes the transaction decision to the participants and their replicas. Once the client's decision is received by more than a replica quorum of any participant, HACommit will guarantee that the transaction is committed or aborted according to the client's decision despite failures of the client or the participant replicas. Therefore, the client can safely end the transaction once it has received acknowledgement from a replica quorum of any participant. The transaction will be committed at all participant replicas once they receive the client's commit decision. An example of transaction processing using HACommit is illustrated in Figure~\ref{fig:hacommit}.

\section{Processing the Last Operation}%
\label{sec:lastop}

On processing the last operation of the transaction, the client sends the last operation to participants holding relevant data, indicating about the last operation. For other participants, the client sends an empty operation as the last operation. All participants process the last operation--those receiving an empty operation does no processing. They check locally whether a commit for the transaction can violate any ACID property and vote accordingly. They replicate their votes and the transaction context to their replicas respectively before responding to the client. The replication of participant votes and the transaction context is required to survive the votes and guarantee voting consistency in case of participant failures. The participants piggyback their votes on their response to the client's last operation request after the replication. The client makes its decision on receiving responses from all participants.

\textbf{Transaction context}. The transaction context must include the \emph{transaction ID} and the \emph{shard IDs}. The transaction ID uniquely identifies the transaction and distinguishes the Paxos instance for the commit process. The shard IDs are necessary to compute the set of participant IDs, which constitute the configuration information of the Paxos instance for commit. This configuration information must be known to all acceptors of Paxos.

In case when inconsistent replication~\cite{tapir} is used in operation processing, the transaction context must also include \emph{relevant writes}.  Relevant writes are writes operating on data held by a participant and its replicas. The relevant writes are necessary in case of participant failures. With inconsistent replication, participant replicas might not process the same writes for a transaction as the participant. Consider when a set of relevant writes are known to the participant but not its replicas. The client might fail after sending the Commit decision to participants. In the mean time, a participant fails and one of its replica acts as the new participant. Then, the recovery proposers propose the same Commit decision. In such case, the new participant will not know what writes to apply when committing the transaction. To reduce the data kept in the transaction context, the relevant writes can be recorded as commands~\cite{cmdlogging}.

\section{The Commit Process}%
\label{sec:commit}

In HACommit, the client commits or aborts a transaction by initiating a Paxos instance.

\subsection{Background: the Paxos Algorithm}

A run of the Paxos algorithm is called an instance. A Paxos instance reaches a single consensus among the participants. An instance proceeds in rounds. Each round has a ballot with a unique number \emph{bid}. Any would-be proposer can start a new round on any (apparent) failure. Each round generally consists of two phases~\cite{livepaxos} (\emph{phase-1} and \emph{phase-2}), and each phase involves one communication roundtrip. The consensus is reached when one active proposer successfully finishes one round. Participants in the consensus problem are generally called acceptors in Paxos. However, in an instance of Paxos, if a proposer is the \emph{only} and the \emph{initial} proposer, it can propose any value to be accepted by participants as the consensus, incurring one communication roundtrip between the proposer and the participants~\cite{paxosSimple}.

Paxos is commonly used in reaching the consensus among a set of replicas. Each Paxos instance has a configuration, which includes the set of acceptors and learners. Widely used in reaching replica consensus, Paxos is generally used with its configuration staying the same across instances~\cite{megastore}. The configuration information must be known to all proposers, acceptors and learners. Take data replication for example. The set of data replicas are acceptors and learners. The leader replica is the initial proposer and all other replicas are would-be proposers. Clients send their writing requests to the leader replica, which picks one write or a write sequence as its proposal. Then the leader replica starts a Paxos instance to propose its proposal to the acceptors. In practice, the configuration can stays the same across different Paxos instances, e.g., writes to the same data at different time.

\subsection{The One-Phase Commit Process}

In HACommit, the client is the \emph{only} and the \emph{initial} proposer of the Paxos instance, as each transaction has a unique client. As a result, the client can commit the transaction in one communication roundtrip to the participants.

The commit process starts from the second phase (phase-2) of the Paxos algorithm. That is, the client first sends a phase-2 message to all participants. To guarantee the correctness, the exploitation of the Paxos algorithm must strictly comply with the algorithm specification. Complying with the Paxos algorithm, the phase-2 message includes a ballot number \emph{bid}, which is equal to zero, and the proposal for commit, which can be \emph{commit} or \emph{abort}. On receiving the phase-2 message, a participant records the ballot number and the outcome for the transaction locally. Then it commits the transaction by applying the writes and releasing all data items; or, it aborts the transaction by rolling back the transaction and releasing all data items. In the mean time, the participant invokes the replication layer to replicate the result to its replicas. Afterwards, each participant acknowledges the client. Alternatively, the client can send the phase-2 message to all participants and their replicas. Each participant replica also follows the same processing procedure as its participant's. Then, the client waits responses from all participants and their replicas.

\subsection{Participant Acknowledgements}

For any participant, if the acknowledgements by a quorum of its replicas are received by the client, the client can safely end the transaction. In fact, the commit process is not finished until all participants acknowledge the client. But any participant failing to acknowledge can go through the failure recovery process (Section~\ref{sec:failure}) to successfully finish the commit process. In HACommit, all participants must finally acknowledge the acceptance of the client's proposal so that the transaction is committed at all data operated by the transaction.

The requirement for participants' acknowledgements is different from that for the quorum acceptance in the original Paxos algorithm. In Paxos, the Consensus is reached if a proposal is accepted by more than a quorum of participants. The original Paxos algorithm can tolerate the failures of both participants (acceptors) and proposers. HACommit uses the client as the initial proposer and the participants as acceptors and would-be proposers when exploiting Paxos for the commit process. In its Paxos exploitation, HACommit only tolerates the failures of the initial proposer and would-be proposers. However, the failure of participants (i.e., acceptors) can be tolerated by the participant replication, which can also exploit consensus algorithms like Paxos.

\subsection{Distinguishing Concurrent Commits}

Each Paxos instance corresponds to the commit of one transaction, but one participant can engage in multiple Paxos instances for commit, as the participant can involve in multiple concurrent transactions. To distinguish different transactions, we include a transaction ID in the phase-2 message, as well as in all messages sent between clients and participants. A transaction \emph{T} is uniquely identified in the system by its ID \emph{tid}, which can be generated using distributed methods, e.g., UUID~\cite{uuid}.

\subsection{Paxos Configuration Information}
\label{sec:configupdate}

Different from those Paxos exploitations where the configuration stay the same across multiple instances, HACommit has different configurations in Paxos instances for different transaction commits. The set of participants is the configuration of a Paxos instance. Each transaction has different participants, leading to different configurations of Paxos instances for commit. As required by the algorithm, the configuration must be known to all proposers and within the configuration. A replacing proposer (i.e., a recovery node) needs the configuration information to continue the algorithm after the failure of a previous proposer. The first proposer of the commit instance is the transaction client, which is the only node with complete information of the configuration. If the client fails, the configuration information might get lost. In fact, a client might fail before the transaction comes to the commit step. Then a replacing proposer will hardly have enough configuration to abort the dangling transaction.

To guarantee the availability of the configuration information, we include the configuration information in the phase-2 message. Besides, as the configuration is expanding and updating after a new operation is processed, the client must send the up-to-date configuration to \emph{all} participants contacted so far on processing each operation. In case that a participant fails and one of its replicas take its place, the configuration must be updated and sent to all replicas of all participants. The exact configuration of the Paxos instance for commit will be formed right on the processing of the last transactional operation. In this way, each participant replica keeps locally an up-to-date copy of the configuration information. As a participant can fail and be replaced by its replicas, HACommit does not rely on participant IDs for the configuration reference. Instead, it records the IDs of all shards operated by the transaction. With the set of shard IDs, any server in the system shall find out the contemporary set of participants easily.

\section{Failure Recovery}%
\label{sec:failure}

In the design of HACommit, we assume that, if a client or a participant replica fails, it can only fail by crashing. In the following, we describes the recovery mechanisms for client failure and participant replica failure respectively.

\subsection{On Client Failure}

In HACommit, all participants are all candidates of recovering nodes for a failure. We call recovering nodes as \emph{recovery proposers}, which act as would-be proposers of the commit process. The recovery proposers will be activated on client failure. In an asynchronous system, there is no way to be sure about whether a client actually fails. In practical implementations, a participant can keep a timer on the duration since it has received a message from the current proposer. If the duration has exceeded a threshold, the participant considers the current proposer as failed. Then it considers itself as the recovery proposer.

A recovery proposer must run the complete Paxos algorithm to reach the consensus \emph{safely} among the participants. As any would-be proposer can start a new round on any (apparent) failure, multiple rounds, phases and communications roundtrips will be involved on client failures.

Although complicated situations can happen,  the participants of a transaction will reach the same outcome eventually, if they ever reach a consensus and the transaction ends. For example, as delayed messages cannot be distinguished from failures in an asynchronous system, the current proposer might in fact have not failed. Instead, its last message has not reached a participant, which considers the proposer as failed. Or, multiple participants considers the current proposer as failed and starts a new round of Paxos simultaneously. All these situations will not impair the safety of the Paxos algorithm~\cite{paxos}.

\subsubsection{The Recovery Process}

A recovery proposer starts the recovery process by starting a new round of the Paxos instance from the first phase. In the first phase, the new proposer will update the ballot number $bid$ to be larger than any one it has seen. It sends a phase-1 message with the new ballot number to all participants. On receiving the phase-1 message with $bid$, if a participant has never received any phase-1 message with ballot number greater than $bid$, it responds to the proposer. The response includes the accepted transaction decision and the ballot number on which the acceptance is made, if the participant has ever accepted any transaction decision.

If the proposer has received responses to its phase-1 message from all participants, it sends a phase-2 message to all participants. The phase-2 message has the same ballot number as the proposer's last phase-1 message. Besides, the transaction outcome with the highest ballot number in the responses is proposed as the final transaction outcome; or, if no accepted transaction outcome is included in responses to the phase-1 message, the proposer proposes ABORT to satisfy the assumptions of the CAC problem. Unless the participant has already responded to a phase-1 message having a ballot number greater than $bid$, a participant accepts the transaction outcome and ends the transaction after receiving the phase-2 message. The participant acknowledges the proposer accordingly. After receiving acknowledgements from all participants, the new proposer can safely end the transaction.

\subsubsection{Liveness}%
\label{sec:liveness}

To guarantee liveness, HACommit adopts the assumption commonly made for Paxos. That is, one proposer will finally succeed in finishing one round of the algorithm. In HACommit, if all participants consider the current proposer as failed and starts a new round of Paxos simultaneously, a racing condition among new proposers could be formed in the first phase of Paxos. No proposer might be able to succeed in finishing the second phase of Paxos, making the liveness of commit not guaranteed. Though rarely happening, the racing condition among would-be proposers must be avoided in Paxos~\cite{paxos} for the liveness consideration. In actual implementations, the random back-off of candidates is enough to resolve the racing situation~\cite{megastore,livepaxos}; or, some leader election~\cite{livepaxos} or failure detection~\cite{failureDetector} services outside the algorithm implementation might be used.

\subsection{On Participant Replica Failures}

HACommit can tolerate not only client failures, but also participant replica failures. It can guarantee continuous data availability if more than a quorum of replicas are accessible for each participant in a transaction. In case that quorum replica availability cannot be guaranteed, HACommit can be blocked but the correctness of atomic commit is guaranteed anyhow~\cite{paxos}. The high availability of data enables a recovery process based on \emph{replicas} instead of logging, though logging and other mechanisms like checkpointing~\cite{hstore} and asynchronous logging~\cite{cmdlogging} can fasten the recovery process.

Failed participant replicas can be recovered by copying data from the correct replicas of the same participant. Or, recovery techniques used in consensus and replication services~\cite{chubby,remus} can be employed for the replica recovery of participants. Although one replica is selected as the leader (i.e., the participant), the leader replica can easily be replaced by other replicas of the same participant~\cite{chubby}. If a participant failed before sending its vote to its replicas, the new leader will make a new decision for the vote. Otherwise, as the vote of a participant is replicated before sending to the coordinator, this vote can be kept consistent during the change of leaders. Besides, the client hast sent the transaction outcome to all participants and their replicas in the commit process. Thus, failed participant replicas can be recovered correctly as long as the number of failed replicas for a participant is tolerable by the consensus algorithm in use.

We assume there are fewer failed replicas for each participant than that is tolerable by the highly-available datastore. This is generally satisfied, as the number of replicas can be increased to tolerate more failures. If unfortunately the assumption is not met, the participant without enough replicas will not respond to the client so as to guarantee replica consistency and correctness. The commit process will have to be paused until all participants have enough active replicas. Though not meeting the assumption can impair the liveness of the protocol, HACommit can guarantee the correctness of commit and the consistency of data anyhow.

\section{Evaluation}%
\label{sec:eval}%

Our evaluation explores three aspects: (1) commit performance of HACommit---it has smaller commit latency than other protocols and this advantage increases as the number of participants per transaction increases; (2) fault tolerance of HACommit---it can tolerate client failures, as well as server failures; and, (3) transaction processing performance of HACommit---it has higher throughputs and lower average latencies than other protocols.

\subsection{Experimental Setup}

We compare HACommit with two-phase commit (2PC), replicated commit (RCommit)~\cite{rcommit} and MDCC~\cite{mdcc}. Two-phase commit (2PC) is still considered the standard protocol for committing distributed transactions. It assumes no replication and is not resilient to single node failures. RCommit and MDCC are state-of-the-art commit protocols for distributed transactions over replicated as HACommit is. It has better performance than the approach that layers 2PC over the Paxos algorithm~\cite{megastore,spanner}. MDCC guarantee only isolation levels weaker than serializability. The same concurrency control scheme and the same storage management component are used for HACommit, 2PC and RCommit. These three implementations use the consistency level of serializability. Compared to the implementations for 2PC and RCommit, the HACommit implementation also supports the weak isolation level of read committed~\cite{critique}. The evaluation of MDCC is based on its open sources~\cite{mdcccode}.

We evaluate all implementations using the Amazon EC2 cloud. We evaluate each implementation using a YCSB-based benchmark~\cite{ycsb}. As our database completely resides in memory and the network communication plays an important role, we deploy the systems over memory-optimized instances of r3.2xlarge (with 8 cores, 60GB memory and high-speed network). Unless noted otherwise, all implementations are deployed over eight nodes. The cross-node communication roundtrip is about 0.1 milliseconds.

For HACommit, RCommit and MDCC, the database is deployed with three replicas. For 2PC, no replication is used. Generally, 2PC requires buffer management for durability. We do not include one for 2PC and in-memory database is used instead. The durability is guaranteed through operation logging. As buffer management takes up about one fifth of the local processing time of transactions, our 2PC implementation without buffer management should perform faster than a typical 2PC implementation.

In all experiments, each server runs a server program and a test client program. By default, each client runs with 10 threads. Each data point in the graphs represents the median of at least five trials. Each trial is run for over 120s with the first and last quarter of each trial elided to avoid start up and cool down artifacts. For all experimental runs, clients recorded throughput and response times. We report the average of three sixty-second trials.

In all experiments, we preload a database containing a single table with 10 million records. Each record has a single primary key column and 1 additional column each with 10 bytes of randomly generated string data. We use small-size records to focus our attention on the key performance factors. Accesses to records are uniformly distributed over the whole database. In all workloads, transactions are committed if no data conflicts exist. That is, all transaction aborts in the experiments are due to concurrency control requirements.
\begin{figure}[!t]
      \centering
      \includegraphics[width=0.45\textwidth]{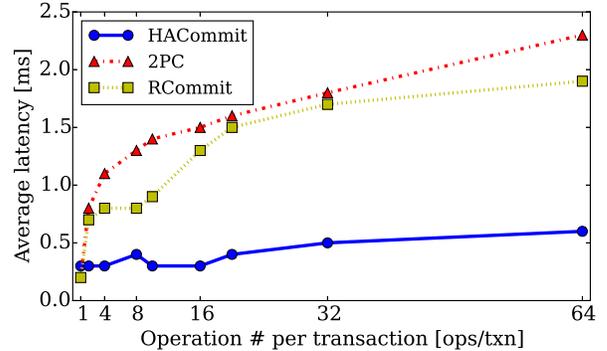}
      \caption{Commit latencies when increasing the number of operations per transaction.}\vspace{-12pt}
      \label{fig:commitLatency} 
\end{figure}%
\begin{figure*}[t]
  \begin{minipage}{0.31\textwidth}
      \centering
       \includegraphics[width=\textwidth]{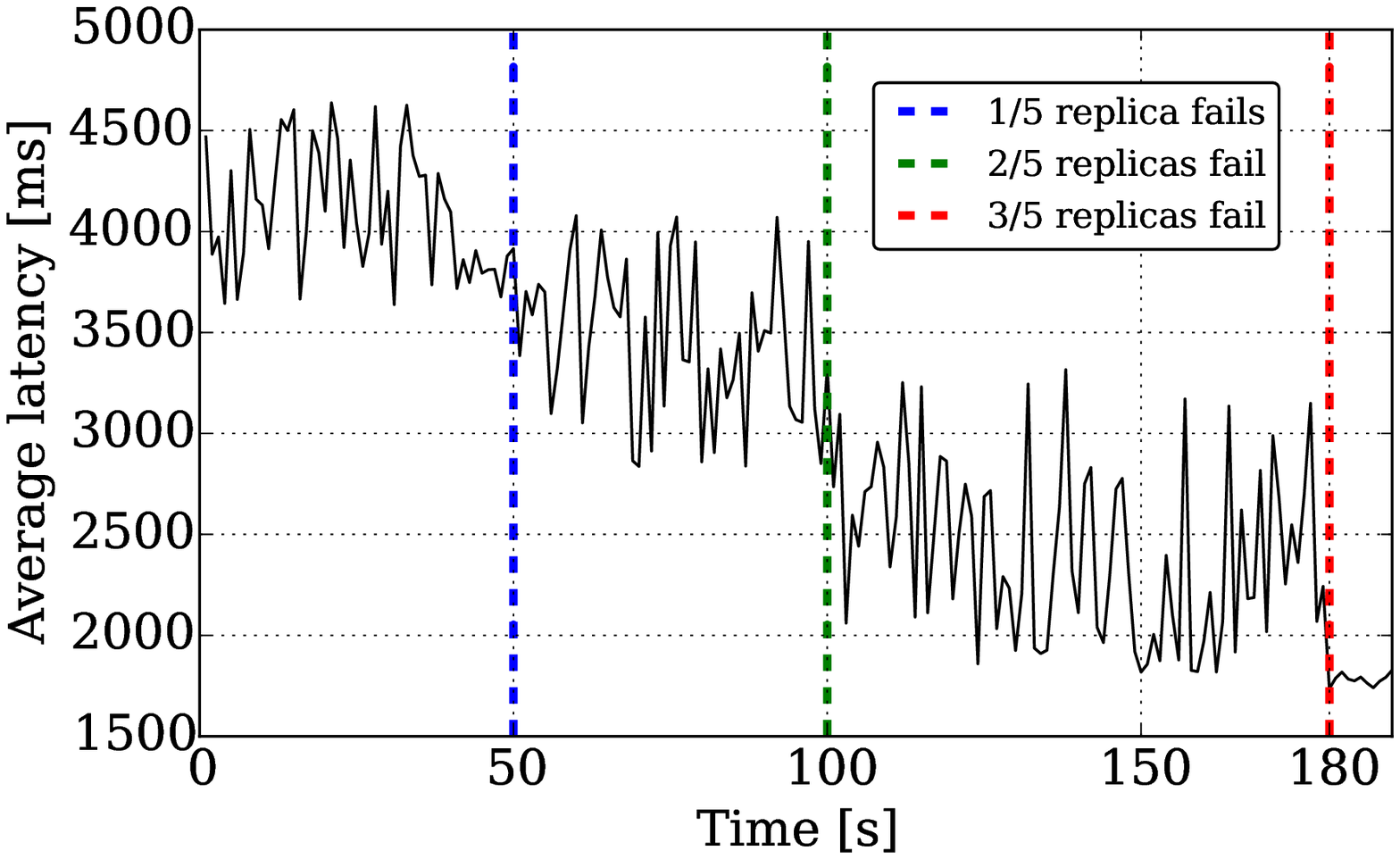}
      \caption{Transaction latency variations during server failures.}\vspace{-12pt}
      \label{fig:partFailLatency} 
  \end{minipage}
  \hspace{2pt}
 \begin{minipage}{0.31\textwidth}
      \centering
      \includegraphics[width=\textwidth]{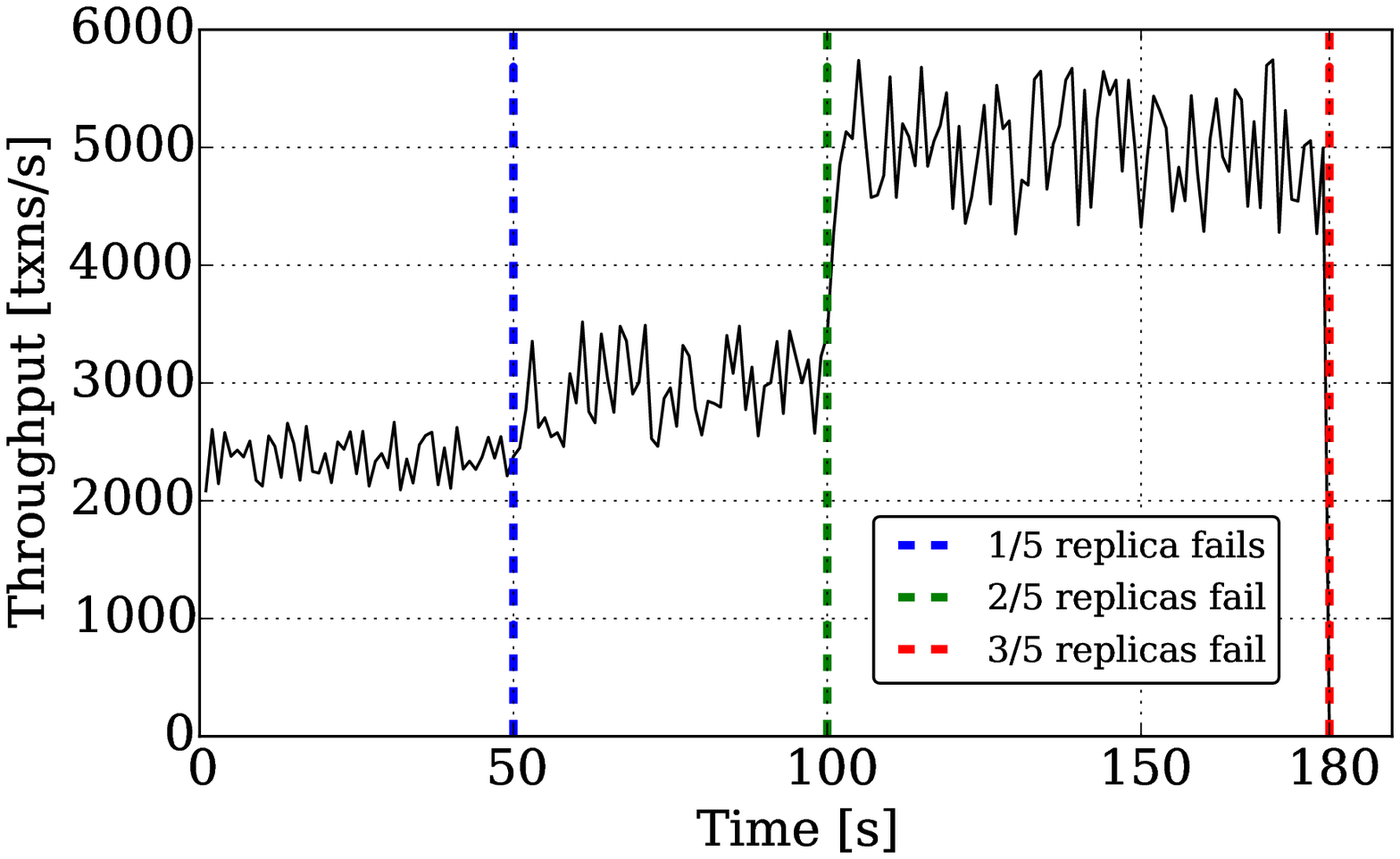}
      \caption{Transaction throughput variations during server failures.}\vspace{-12pt}
      \label{fig:partFailThroughput} 
  \end{minipage}
 \hspace{2pt}
 \begin{minipage}{0.37\textwidth}
 \vspace{3pt}
     \centering
     \includegraphics[width=\textwidth]{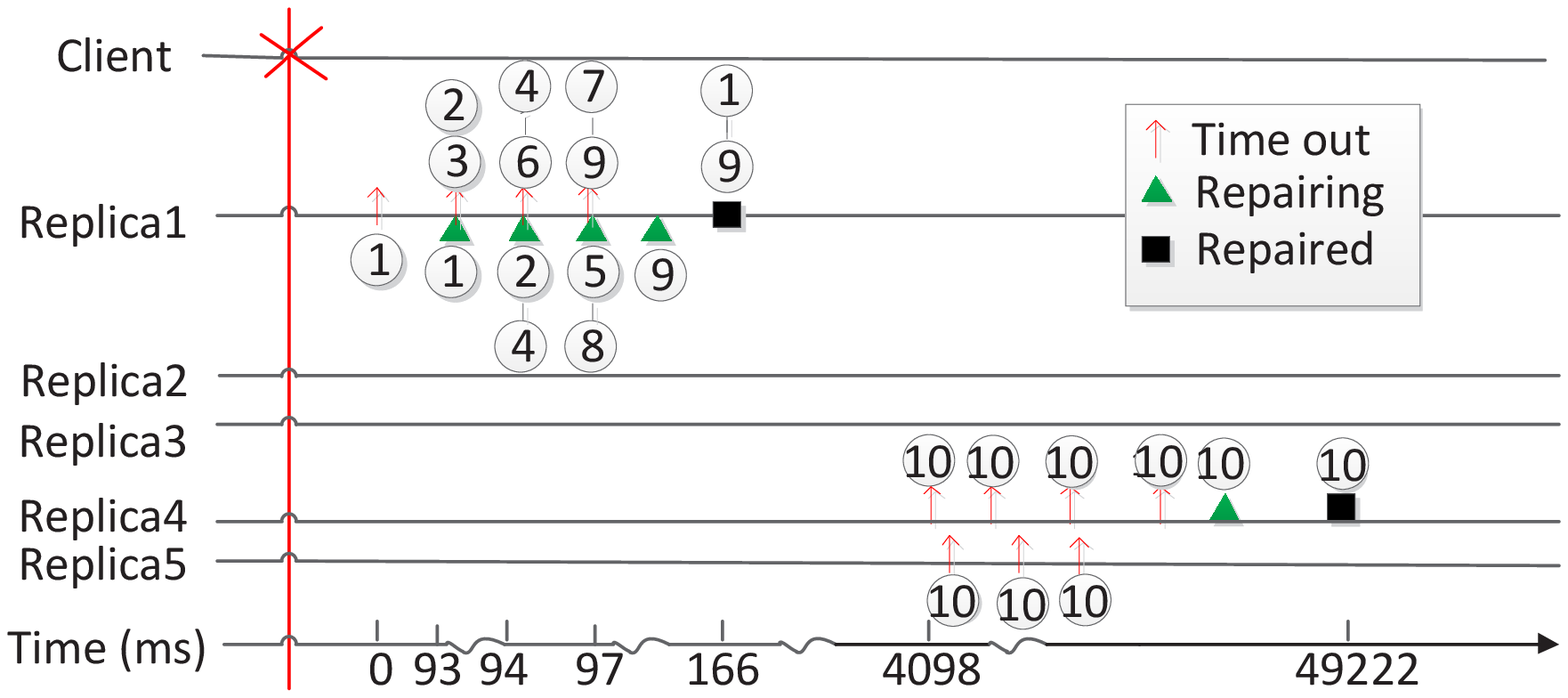}\vspace{6pt}
      \caption{HACommit's behavior on a client failure (circled numbers are transactions).}\vspace{-12pt}
      \label{fig:clientFailure} 
    \end{minipage}
\end{figure*}
\subsection{Commit Performance}

As we are targeting at transaction commit protocols, we first examine the actual costs of the commit process. We study the duration of the commit process. We do not compare the commit process of HACommit with that of MDCC because the latter integrates a concurrency control process; comparing only the commit process of the two protocols is unfair to MDCC.

HACommit outperforms 2PC and RCommit in varied workloads. Figure \ref{fig:commitLatency} shows the latencies of commit. We vary the number of operations per transaction from 1 to 64. The advantage of HACommit increases as the number of operations per transaction increases. When a transaction has 64 operations, HACommit can commit in one fifth of the time 2PC does. This performance is more significant as it seems, as HACommit uses replication and 2PC does not. That means, HACommit has $n-1$ times more participants than 2PC in the commit, where $n$ is the number of replicas.

HACommit's commit latency increases slightly as the number of operations increases to 20. On committing a transaction, the system must apply all writes and release all locks. When the number of operations is small, applying writes and releasing locks in the in-memory database cost a small amount of time, as compared to the network communication roundtrip time (RTT). As the number of operations increases, the time needed increases slightly for applying all writes in the in-memory database. Accordingly, the commit latency of HACommit increases.

2PC and RCommit have increased commit latencies when the number of operations per transaction increases. They need to log new writes for commit and the old values of data items for rollback, thus the time needed for the prepare phase increases as the number of writes goes up, leading to a longer commit process. 2PC has a higher commit latency than RCommit, because in our implementations, 2PC must log in-memory data and RCommit relies on replication for fault tolerance.

\subsection{Fault-Tolerance}

In the fault-tolerance tests, we examine the behaviors of HACommit under both client failures and server failures. The evaluation result demonstrates that no transaction is blocked under server failures and the client failure, as long as a quorum of participant replicas are accessible.

We use five replicas and initiate one client  in the fault tolerance tests. To simulate failures, we actively kill a process in the experiments. The network module of our implementations can instantly return an error in such case. Our implementation processes the error as if connection timeout on node failures happens.

Figure \ref{fig:partFailLatency} shows the evolution of the average transaction latency in a five replica setup that experiences the failure of one replica at 50, 100 and 180 seconds respectively. The corresponding throughputs are shown in Figure \ref{fig:partFailThroughput}. The latencies and throughputs are captured for every second. At 50 and 100 seconds, the average transaction latency decreases and the throughput increases. With PCC, reads in the HACommit implementation take up a great portion of time. The failure of one replica means that the system can process fewer reads. Hence, this leads to lower average latencies and higher throughputs for read transactions, as well as for all transactions. At 180 seconds, we failed one more replicas, violating the quorum availability assumption of HACommit. The thoughput drops to zero immediately because no operation or commit process can succeed at all. The HACommit implementation uses timeouts to detect failures and quorum reads/writes. As long as a quorum of replicas are available for every data item, HACommit can process transactions normally.

We also examine how HACommit behaves under transaction client failures. We deliberately kill the client in an experiment. Each server program periodically checks its local transaction contexts to see if any last contact time exceeds a timeout period. We set the timeout period to be 15 seconds. Figure \ref{fig:clientFailure} visualizes the logs on client failures and demonstrates how participants recover from the client failure.

In Figure \ref{fig:clientFailure}, replicas represent participants. The cross at the client line represents the failure of the client. The circled numbers represent unended transactions. The time axis at the bottom stretches from left to right. The moment when the first transaction is detected to be unended is denoted as the beginning of the time axis. A transaction is named on the time that it is discovered to be unended, i.e., transaction 1 is the first transaction to be detected not to be unended. A replica can detect a transaction is unended because there are timeouts on the last time when a processing message is received at the replica. Timeouts are specified by arrows.

Replica 1 has the smallest node ID. It detects unended transactions 1 to 9 and starts pushing them to an end through a repairing process. We have synchronized the clocks of nodes. For simplicity, we use the moment when replica 1 detects transaction 1 has a last contact time exceeding the timeout period as the beginning of the time axis. It takes about 100 milliseconds for replica 1 to repair each transaction. Replica 1 aborts the nine transactions in the repairing process because no transaction outcome has ever been accepted by any replica. The transaction 10 is later detected by replica 4. However, replica 4 waits for four timeout periods before it actually initiates a repairing process for the transaction. The reason is that transaction 10 has been committed at replica 1, 2 and 3. Replica 4 finally commits the transaction. Replica 5 also detects transaction 10, but it does not initiate any repairing process. Before replica 5 starts repairing transaction 10, the transaction is already committed in the repairing process initiated by replica 4.
\begin{figure*}[t]
  \begin{minipage}{0.33\textwidth}
      \centering
      \includegraphics[width=\textwidth]{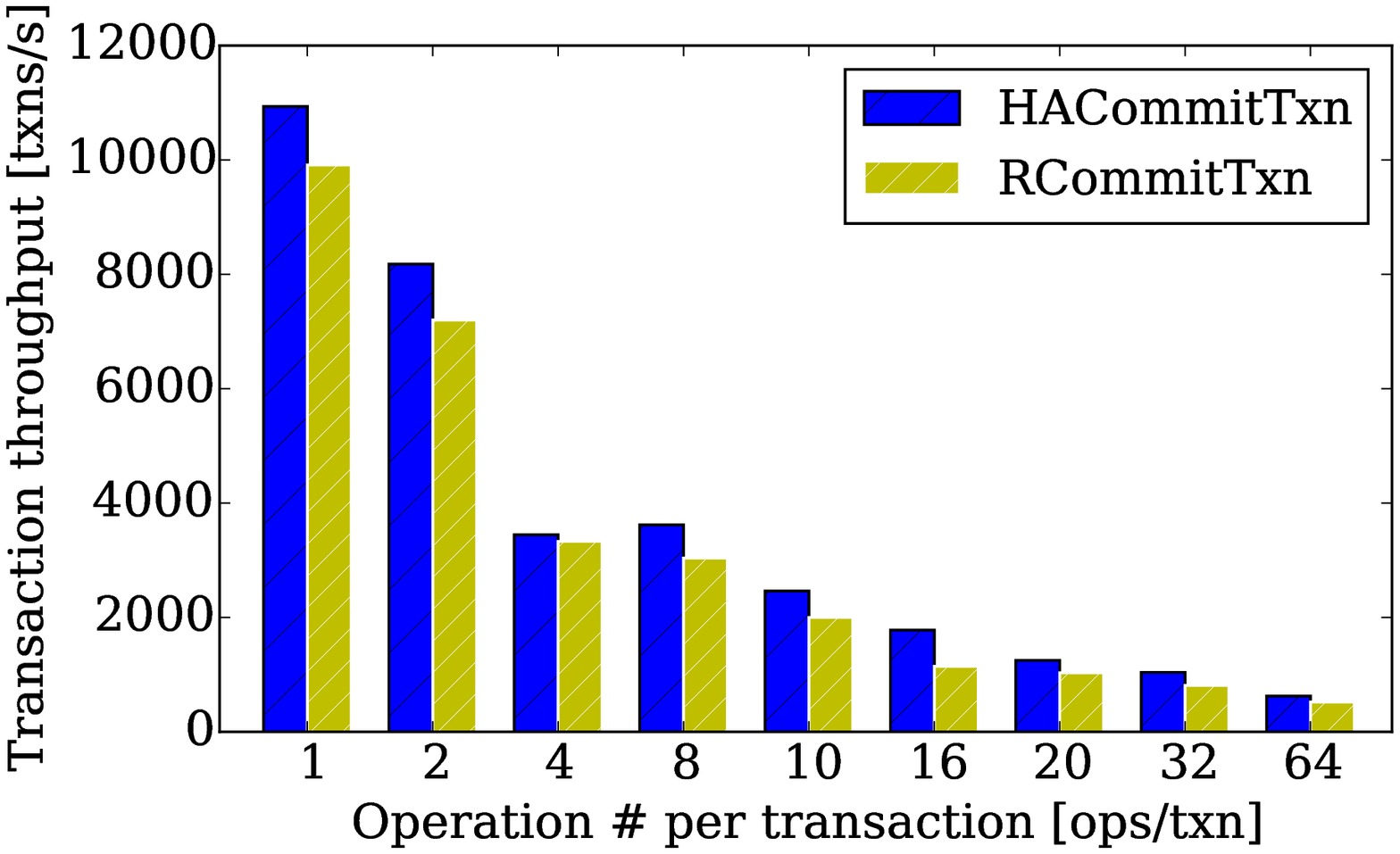}
      \caption{Transaction throughput: HACommit vs. RCommit.}
      \label{fig:txnThroughput} 
  \end{minipage}
  \hspace{2pt}
 \begin{minipage}{0.33\textwidth}
      \centering
       \includegraphics[width=\textwidth]{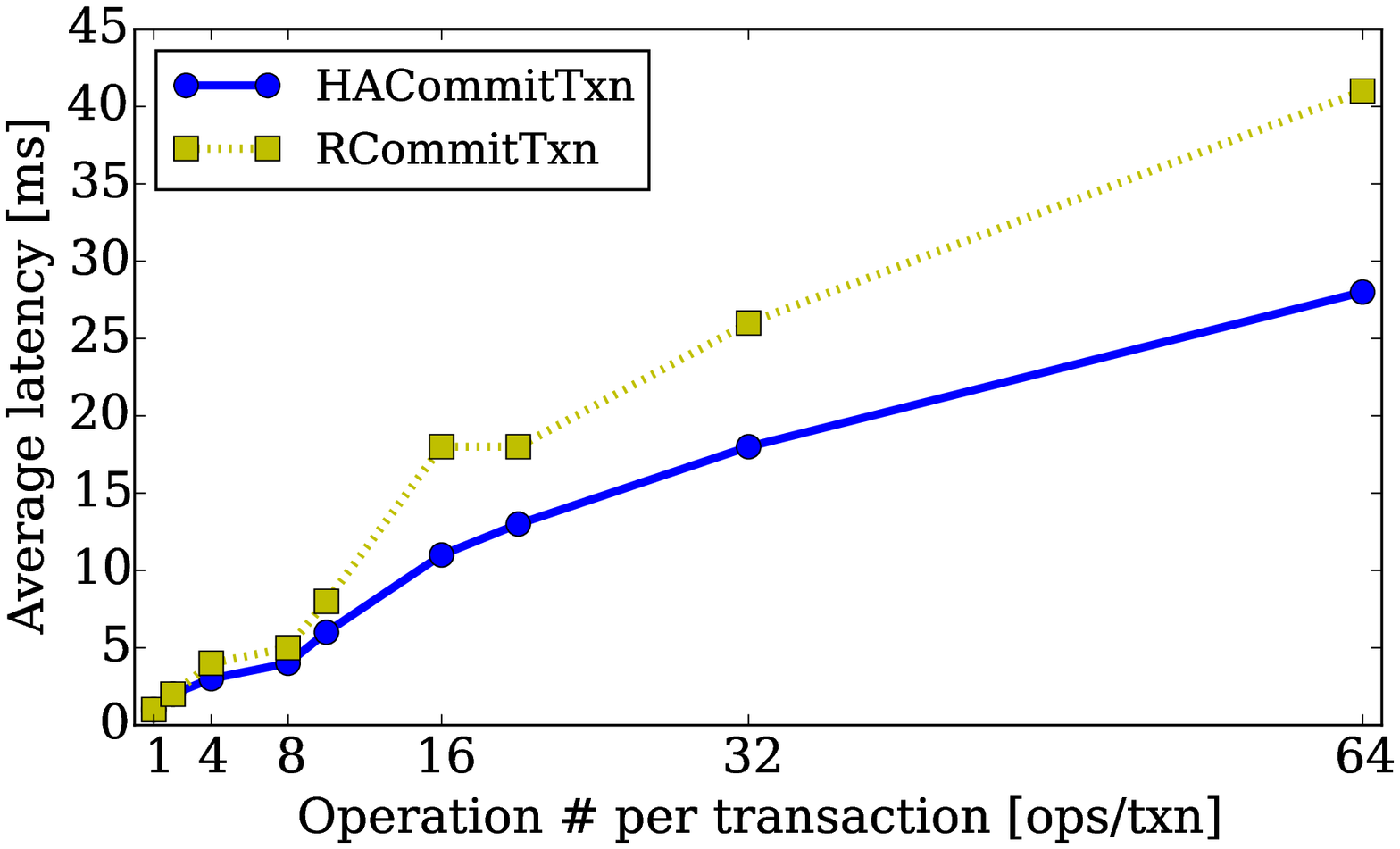}
      \caption{Transaction average latency: HACommit vs. RCommit.}
      \label{fig:txnLatency} 
  \end{minipage}
 \hspace{2pt}
 \begin{minipage}{0.33\textwidth}
     \centering
     \includegraphics[width=\textwidth]{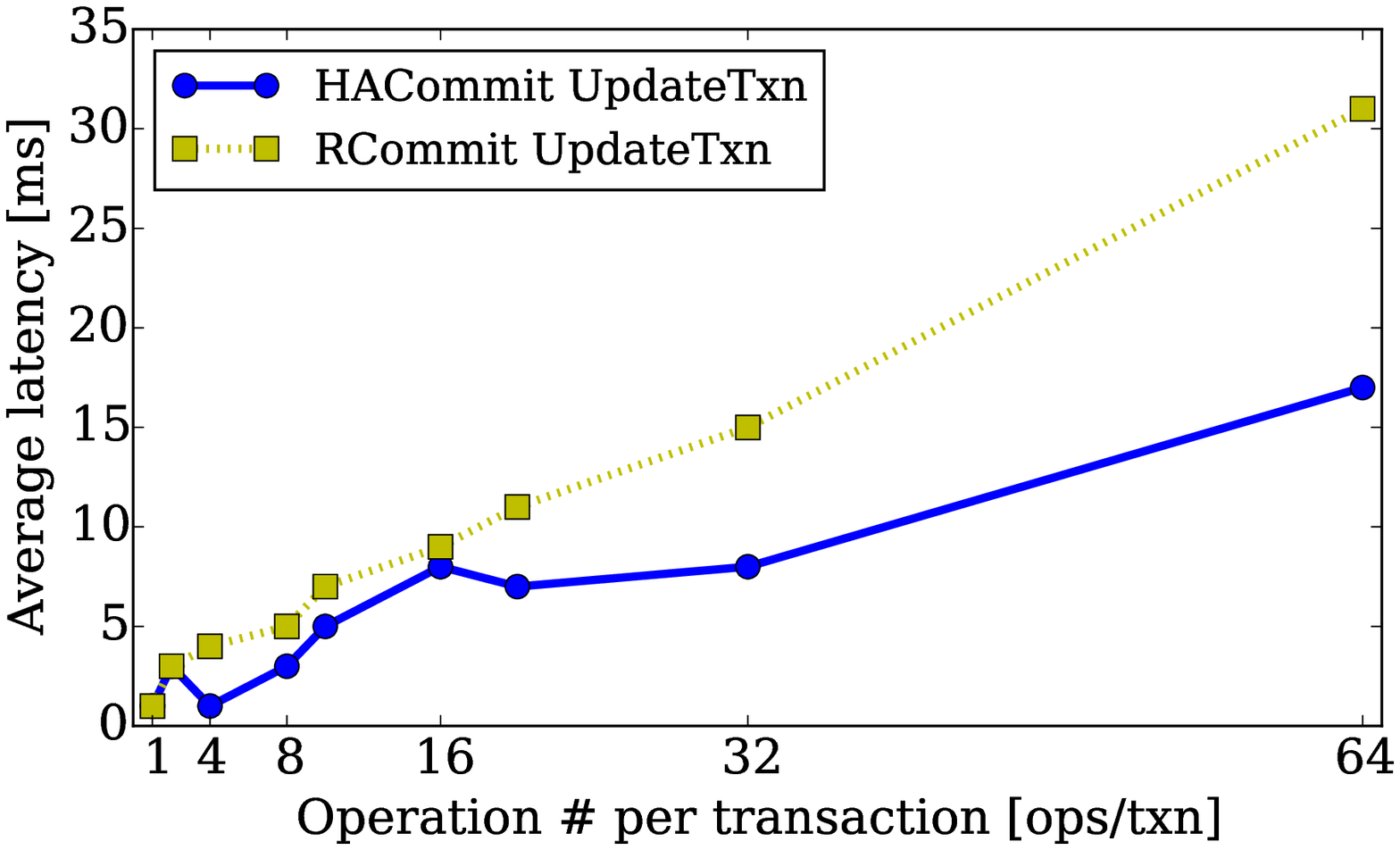}
      \caption{Latency of update transaction: HACommit vs. RCommit.}
      \label{fig:updatetxnLatency} 
    \end{minipage}
\end{figure*}
\begin{figure*}[t]
  \begin{minipage}{0.33\textwidth}
      \centering
      \includegraphics[width=\textwidth]{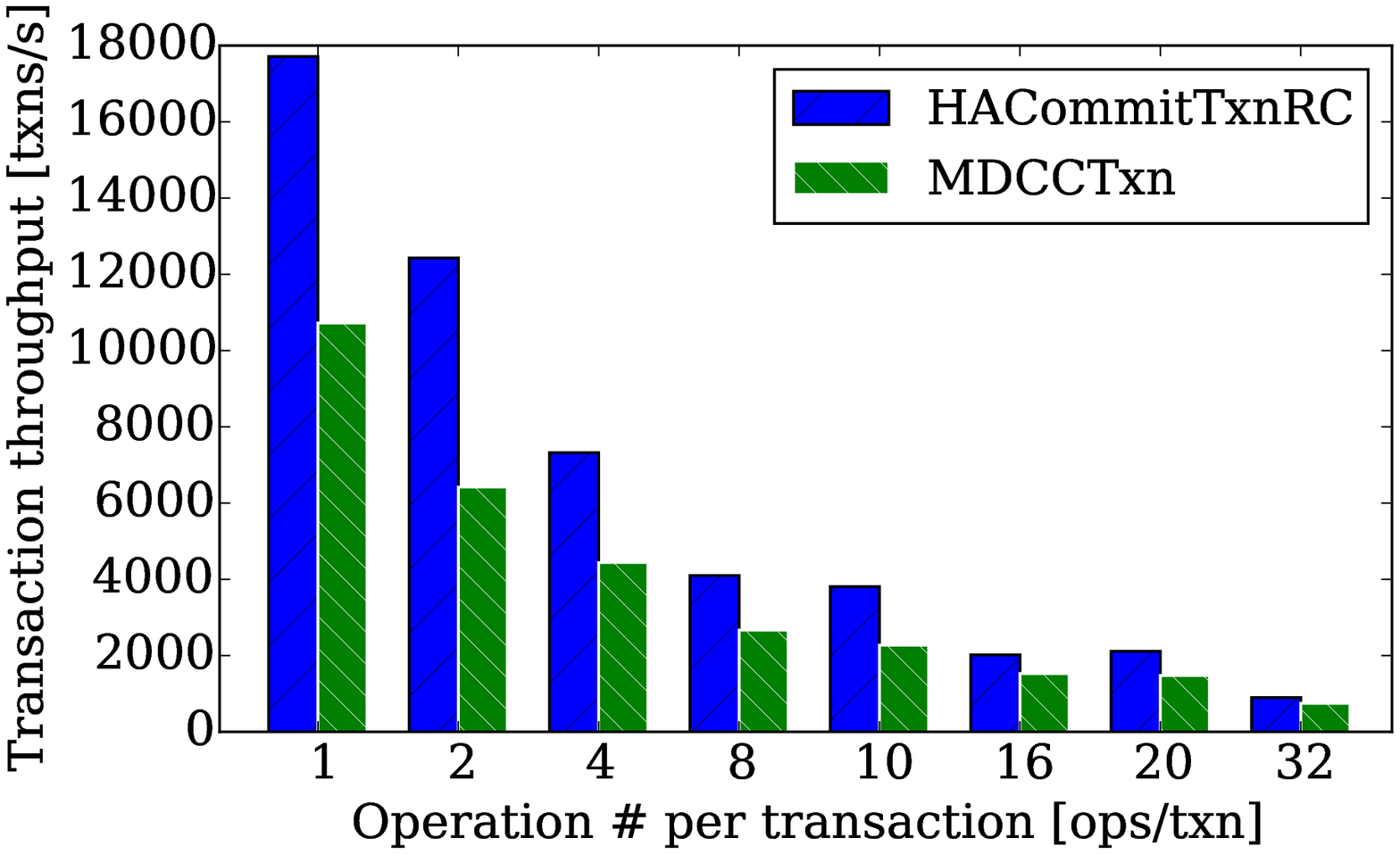}
      \caption{Transaction throughput under read-committed CC: HACommit vs. MDCC.}
      \label{fig:isoThroughput} 
  \end{minipage}
  \hspace{2pt}
 \begin{minipage}{0.33\textwidth}
      \centering
      \includegraphics[width=\textwidth]{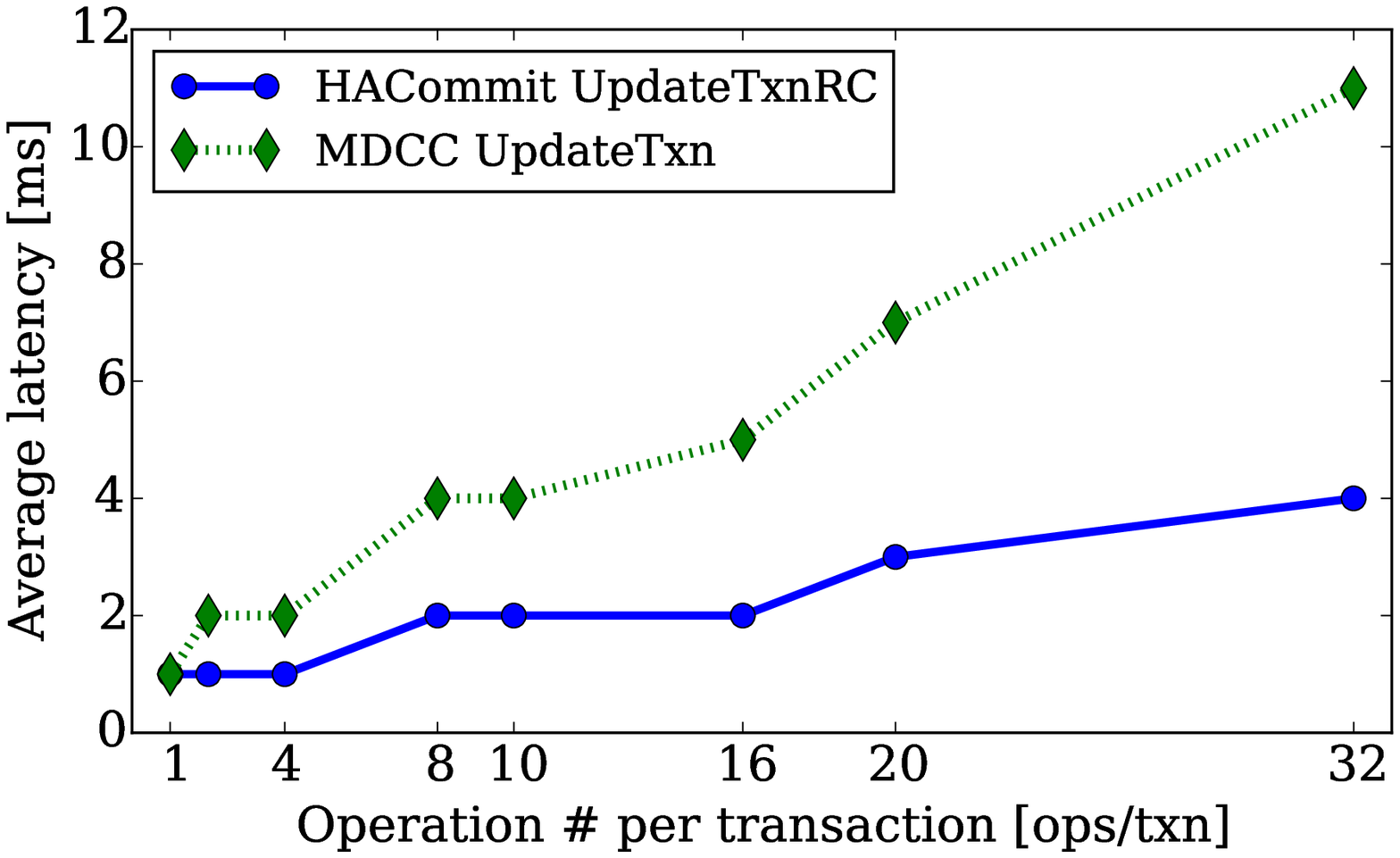}
      \caption{Latency of UPDATE transactions under read-committed CC: HACommit vs. MDCC.}
      \label{fig:isoWriteLatency} 
  \end{minipage}
 \hspace{2pt}
 \begin{minipage}{0.33\textwidth}
     \centering
      \includegraphics[width=\textwidth]{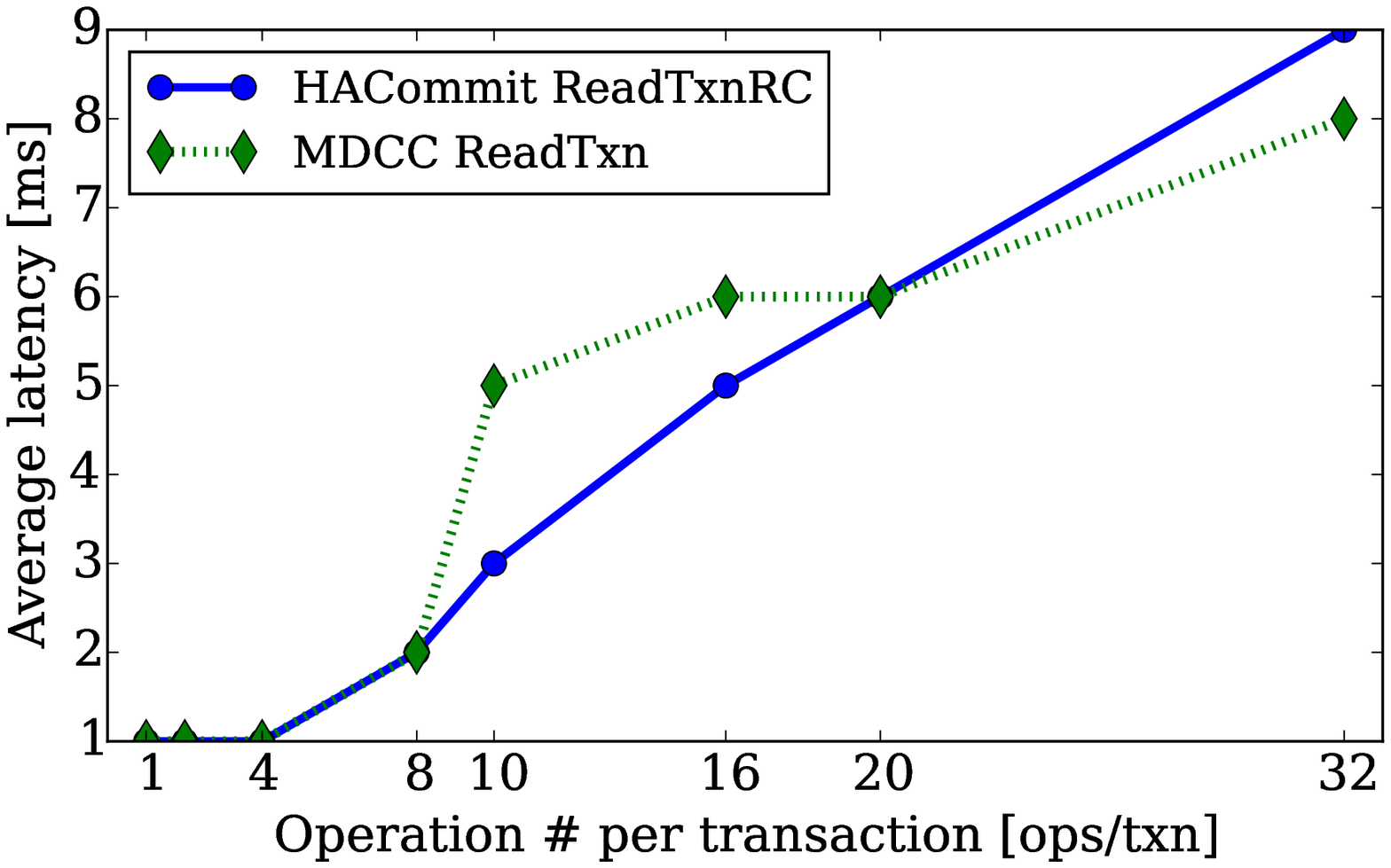}
      \caption{Latency of READ transactions under read-committed CC: HACommit vs. MDCC.}
      \label{fig:isoReadLatency} 
    \end{minipage}
\end{figure*}
\subsection{Transaction Throughput and Latency}

We evaluate the transaction throughput and latency when using different commit protocols. In the experiments, on the failure of lock acquisition, we retry the same transaction until it successfully commits. Each retry is made after a random amount of time.

Figure \ref{fig:txnThroughput} shows the transaction throughputs when using HACommit and RCommit, and Figure \ref{fig:txnLatency} demonstrates the average transaction latencies. The HACommit implementation has larger transaction throughputs than the RCommit implementation in all workloads. Besides, HACommit has lower transaction latencies than RCommit in all workloads. HACommit's advantage on transaction latency increases as the number of operations in a transaction increases in the workloads. As both implementations use the same concurrency control and isolation level, factors leading to HACommit's advantage over RCommit are two-fold. First, no costly logging is involved during the commit. Second, no persistence of data is needed.

We compare the update transaction latencies of HACommit and RCommit in Figure \ref{fig:updatetxnLatency}. Both implementations use the same concurrency control scheme and consistency level. We can see that HACommit outperforms RCommit. As the number of operations increases in the workloads, HACommit's advantage also increases. The advantage of HACommit is still due to a commit without logging and data persistence.

We also examine the transaction throughput and latency when using weaker isolation levels with HACommit. In this case, we compare HACommit against MDCC. We implemented HACommit-RC with the read-committed isolation level~\cite{critique}. This is an isolation level comparable to that guaranteed by MDCC. HACommit-RC differs from HACommit in that it acquires no locks on reads. Figure \ref{fig:isoThroughput} shows the transaction throughputs for HACommit-RC and MDCC. The latencies of update transactions and read transactions are shown  in Figure \ref{fig:isoWriteLatency} and Figure \ref{fig:isoReadLatency}. HACommit-RC has larger transaction throughputs than MDCC in all workloads. The latencies of update transactions are lower in the HACommit-RC implementation than in the MDCC implementation, although they have similar performances in read transactions. Both HACommit-RC and MDCC implement read transactions similarly and guarantee the read-committed consistency level. The reason that HACommit-RC has better performance in transaction throughput and update transaction latency is as follows. MDCC uses optimistic concurrency control, which can cause high abort rates under high contention, leading to lower performance than HACommit-RC, which uses pessimistic concurrency control. Besides, MDCC holds data by outstanding options, leading to the same effect of locking in committed transactions.%

\section{Conclusion}%
\label{sec:conclude}

We have proposed HACommit, a logless one-phase commit protocol for highly-available datastores. In contrast to the classic vote-after-decide approach to distributed commit, HACommit adopts the vote-before-decide approach. In HACommit, the procedure for processing the last transaction operation is redesigned to overlap the last operation processing and the voting process. To commit a transaction in one phase, HACommit exploits Paxos and uses the unique client as the initial proposer. To exploit Paxos, HACommit designs a transaction context structure to keep Paxos configuration information. Although client failures can be tolerated by the Paxos exploitation, HACommit designs a recovery process for client failures such that the transaction can actually end with the transaction data visible to other transactions. For participant replica failures, HACommit has participants replicate their votes and the transaction metadata to their replicas; and, a failure recovery process is proposed to exploit the replicated votes and metadata. Our evaluation demonstrates that HACommit outperforms recent atomic commit solutions for highly-available datastores.

\section*{Acknowledgment}

This work is also supported in part by the State Key Development Program for Basic Research of China (Grant No. 2014CB340402) and the National Natural Science Foundation of China (Grant No. 61303054).

\balance

\bibliographystyle{IEEEtran}
\bibliography{IEEEabrv,ref}

\end{document}